\documentclass[prb,floatfix,twocolumn,showpacs,amsmath,amssymb]{revtex4}
\usepackage{graphicx}
\usepackage{dcolumn}
\usepackage{bm}
\begin{document}

\title{Finite compressibility in the low-doping region of the two-dimensional
$t{-}J$ model}
\author{Massimo Lugas, Leonardo Spanu, Federico Becca, and Sandro Sorella}
\affiliation{INFM-Democritos, National Simulation Center and International
School for Advanced Studies (SISSA), I-34014 Trieste, Italy}
\date{\today}

\begin{abstract}
We revisit the important issue of charge fluctuations in the two-dimensional 
$t{-}J$ model by using an improved variational method based on a wave function
that contains both the antiferromagnetic and the d-wave superconducting 
order parameters. In particular, we generalize the wave function introduced 
some time ago by J.P. Bouchaud, A. Georges, and C. Lhuillier 
[J. de Physique {\bf 49}, 553 (1988)] by considering also a {\it long-range} 
spin-spin Jastrow factor, in order to correctly 
reproduce the small-$q$ behavior of the spin fluctuations. We mainly focus
our attention on the physically relevant region $J/t \sim 0.4$ and find that,
contrary to previous variational ansatz, this state is stable against phase 
separation for small hole doping. Moreover, by performing projection 
Monte Carlo methods based on the so-called fixed-node approach, we obtain 
a clear evidence that the $t{-}J$ model does not phase separate 
for $J/t \lesssim 0.7$ and that the compressibility remains finite close to 
the antiferromagnetic insulating state.
\end{abstract}

\maketitle

\section{Introduction}

The possible existence of charge and spin inhomogeneities and their relevance 
for the low-temperature physics of cuprates superconductors 
is a long-standing problem, not yet completely clarified.~\cite{revkivelson}
In particular, the issue is twofold: On one hand, one is interested
to understand the low-energy behavior of microscopic models and the 
possibility to have or not inhomogeneous phases in physically relevant regions;
On the other hand, it is also important to clarify the possible relation 
between charge or spin inhomogeneities and the electronic pairing, which 
may lead to an high critical temperature for superconductivity. 

The original interest in the role of these inhomogeneities dates back to the
works by Emery and Kivelson~\cite{emery,lin} and raised when neutron scattering 
experiments~\cite{tranquada,tranquada2} suggested the possible formation 
of conducting hole-rich regions separated from hole-poor ones with strong
antiferromagnetic moments. Indeed, in most materials, 
the presence of a true phase separation (PS) instability is ruled out by 
the existence of the long-range Coulomb force that prevents the charge to 
accumulate in macroscopic regions,~\cite{jorgensen} only allowing the 
possibility to have a mesoscopic charge segregation, 
i.e., charge density waves (CDW) or the celebrated stripes. 
In the last decade, a great number of direct and indirect evidences for such
charge segregation has been presented in different cuprate and nickelate
materials, stimulating theoretical investigations in simple microscopic 
models.~\cite{revkivelson}
Several authors addressed the possibility of the emergence of PS 
or CDW generating from the competition between the kinetic energy, 
that tends to delocalize the charge carriers, 
and various local interactions (like for instance the on-site Coulomb 
repulsion, the antiferromagnetic super-exchange or the coupling with
some local phonon), that instead tend to freeze the electrons.

Given the complexity of the strongly correlated problem, that contains 
different energy scales, it is very difficult to study its ground-state
and low-energy properties. For instance, by considering mean-field approaches 
it is very easy to overestimate the tendency of charge 
segregation.~\cite{rice,schulz} In this respect,
a great advantage of the variational Monte Carlo technique is that it allows 
one to consider highly correlated wave functions, which are well beyond 
simple mean-field ansatz.~\cite{gros,shiba} 
Then, it would be very important to compare the validity of the ansatz 
considered with exact ground-state properties on fairly large system sizes, 
since the variational approach may fail, especially for low-energy 
properties. This is possible only for bosonic non-frustrated models by means 
of quantum Monte Carlo projection techniques and for fermion systems the 
so-called sign problem prevents one to reach the exact zero-temperature 
properties in a stable way. Nevertheless, very well established 
and efficient approximate approaches are known for fermionic systems, that 
considerably improve the quality of a given variational guess. For instance, 
the so-called fixed node (FN) method (see below for a detailed description 
of the method on lattice models) allows one to obtain the lowest-energy state 
constrained to have the same signs of a given variational wave function.
Therefore, the FN scheme provides a simple procedure to assess the stability 
of a particular variational wave function, its accuracy being related to the 
differences between its properties and the ones obtained with the 
improved FN state. 

In this paper, we will revisit the problem of the PS instability 
in the $t{-}J$ model on the square lattice. This issue has been largely 
considered by several authors in the recent 
past.~\cite{putikka,manousakis,kohno,lee,calandra,putikka2,ivanov} 
Although, a great effort has been done, a general consensus for 
$J/t \lesssim 0.6$ and small hole doping $\delta$ is still laking.
The $t{-}J$ model is defined by:
\begin{equation}\label{hamitj}
{\cal H} = -t \sum_{\langle i,j \rangle \sigma} 
c_{i,\sigma}^{\dagger} c_{j,\sigma} +h.c. +
J \sum_{\langle i,j \rangle} \left ( {\bf S}_i \cdot {\bf S}_j - 
\frac{1}{4} n_i n_j \right ),
\end{equation}
where $\langle \dots \rangle$ indicates the nearest-neighbor sites, 
$c_{i,\sigma}^{\dagger}$ ($c_{i,\sigma}$) creates (destroys) an electron 
with spin $\sigma$ on the site $i$, ${\bf S}_i =(S^x_i,S^y_i,S^z_i)$ is the 
spin operator, $S^\alpha_i = \sum_{\sigma,\sigma^\prime} c_{i,\sigma}^{\dagger} 
\tau^\alpha_{\sigma,\sigma^\prime} c_{i,\sigma^\prime}$, being 
$\tau^\alpha$ the Pauli matrices, and 
$n_i = \sum_{\sigma} c_{i,\sigma}^{\dagger} c_{i,\sigma}$ is the density
operator. We consider a square lattice with $L$ sites and periodic boundary 
conditions rotated by 45 degrees such that $L=2 l^2$, $l$ being an 
odd integer, so that the non-interacting ground state is non-degenerate 
at half filling, thus reducing the finite-size effects. 
Finally, $J$ is the antiferromagnetic exchange constant and 
$t$ the amplitude for nearest-neighbor hopping. In the following we will
take $t=1$.

For very large $J/t$, at small hole doping,  
the ground state is phase separated between undoped regions, with long-range 
antiferromagnetic correlations, and conducting hole-rich regions.
The simple explanation is based on the fact that the magnetic gain in 
accumulating the holes in a given region of space is much larger than the 
loss of the kinetic energy. Therefore, a phase separated state will have a 
lower energy than an homogeneous one. 
By decreasing $J/t$, the situation is much less clear, since the magnetic
gain becomes comparable with the kinetic one.
Emery, Kivelson, and Lin,~\cite{lin} by using simple variational 
arguments, claimed that the ground state of the $t{-}J$ model should phase 
separate for all values of the antiferromagnetic coupling and close to 
half filling. This claim was firstly confirmed by using a more sophisticated 
Monte Carlo technique~\cite{manousakis}, but then disclaimed by other authors,
using slightly different Monte Carlo approaches and series 
expansions.~\cite{kohno,lee,calandra,putikka2}
In particular, two of us showed that, by filtering out the high-energy 
components of a projected BCS wave function, it was possible to obtain an 
homogeneous ground state for $J/t \sim 0.4$.~\cite{calandra}
Later, this approach was questioned in Ref.~\onlinecite{manousakis2}, since 
it was noted that the ground state is still unstable against PS for very 
small hole doping, where our numerical approach had technical problems. 
In particular, it has been shown that Monte Carlo results could indicate an 
instability for $\delta \lesssim 0.05$.
Moreover, it was disappointing that it was not possible to define a stable
variational wave function and that an homogeneous state was obtained only 
after the filtering procedure.
From all the calculations done by different numerical techniques, it is now
clear that, in any case, the $t{-}J$ model for $J/t \sim 0.5$ is on the
verge of charge instabilities, and both PS or CDW can be stabilized with
small perturbations.~\cite{white,white2,beccastripes}

A key issue that was absent in previous calculations and must
be included in a correct description is the presence of
antiferromagnetic correlations at low doping. Recently, by using a 
variational approach that contains both antiferromagnetism and d-wave
pairing, Ivanov~\cite{ivanov} suggested that the antiferromagnetic ordering 
could enhance the instability towards PS. However, in his approach, the 
presence of an antiferromagnetic order parameter in the fermionic determinant
without the presence of a Jastrow term to take into account spin fluctuations 
implies a wrong behavior of the spin properties at small momenta, that in turn 
could also induce incorrect charge properties.
In fact, by using a spin-wave approach for the Heisenberg model, it has been 
shown~\cite{franjio} that an exceptionally accurate description of the 
ground state is obtained by applying a long-range spin Jastrow factor to the
classically ordered state. In the corresponding variational wave function 
it is important that the Gaussian fluctuations induced by the Jastrow 
term are {\it orthogonal} to the direction of the order parameter, in order 
to reproduce correctly the low-energy excitations. A simple generalization 
of this wave function was used to study the Hubbard model at half filling 
and for low doping.~\cite{becca} 
On the other hand, it is well known~\cite{dagotto,randeria,palee} that a 
projected BCS state with $d_{x^2-y^2}$ symmetry and no antiferromagnetic order 
provides an accurate wave function for the low-doping region of the 
$t{-}J$ model and remains rather accurate in energy even at zero doping, 
where a magnetically ordered ground state is well established in two dimensions.
Therefore, in order to have an accurate variational ansatz to describe 
lightly doped correlated insulators, it seems natural to include both 
antiferromagnetic correlations and electronic pairing.~\cite{plekhanov} 

Following these suggestions, we construct a very accurate variational
wave function that describes an energetically stable homogeneous phase.
Moreover, by considering th FN approach, we have a strong evidence in favor
of an homogeneous ground state for $J/t \lesssim 0.7$ for all the accessible 
hole doping.

The paper is organized as follow: in Sec.~\ref{wavefunction} we present 
the improved variational wave function and the FN method, in Sec.~\ref{results}
we show our numerical results, and finally in Sec.~\ref{conclusions} we draw 
our conclusions.  

\section{The variational wave function and the FN method}\label{wavefunction}

In this section we describe the variational state and the generalized FN 
method that is used to filter out its high-energy components.
Our variational ansatz is constructed by applying different projector
operators to a mean-field state:
\begin{equation}\label{wf}
|\Psi_{VMC} \rangle = {\cal J}_s {\cal J}_d {\cal P}_N {\cal P}_G |\Psi_{MF} \rangle,
\end{equation}
where ${\cal P}_G$ is the Gutzwiller projector that forbids double occupied 
sites, ${\cal P}_N$ is the projector onto the subspace with fixed number of 
$N$ particles, ${\cal J}_s$ is a spin Jastrow factor 
\begin{equation}\label{spinjastrow}
{\cal J}_s = \exp \left ( \frac{1}{2} \sum_{i,j} v_{ij} S^z_i S^z_j \right ),
\end{equation}
being $v_{ij}$ variational parameters, and finally ${\cal J}_d$ is a
density Jastrow factor
\begin{equation}\label{densityjastrow}
{\cal J}_d = \exp \left ( \frac{1}{2} \sum_{i,j} u_{ij} n_i n_j \right ),
\end{equation}
being $u_{ij}$ other variational parameters.
The above wave function can be efficiently sampled by standard variational 
Monte Carlo, by employing a random walk of a configuration $|x \rangle$, 
defined by the electron positions and their spin components along the $z$ 
quantization axis. Indeed, in this case, both Jastrow terms are very simple to
compute since they only represent classical weights acting on the 
configuration.

The main difference from previous approaches is the presence of the
spin Jastrow factor and the choice of the mean-field 
state $|\Psi_{MF} \rangle$, that includes both superconducting and 
antiferromagnetic order parameters. Actually, $|\Psi_{MF} \rangle$ is taken 
as the ground state of the mean-field Hamiltonian 
\begin{eqnarray}
\label{meanfield}
&& {\cal H}_{MF} =  
\sum_{i,j,\sigma} t_{i,j} c_{i,\sigma}^{\dagger} c_{j,\sigma} + h.c. 
-\mu \sum_{i,\sigma} n_{i,\sigma}  \nonumber \\
&&+ \sum_{\langle i,j \rangle} \Delta_{i,j} 
(c_{i,\uparrow}^{\dagger} c_{j,\downarrow}^{\dagger} + 
c_{j,\uparrow}^{\dagger} c_{i,\downarrow}^{\dagger} + h.c.) + {\cal H}_{AF},
\end{eqnarray}
where, in addition to the BCS pairing $\Delta_{i,j}$ (with $d$-wave symmetry),
we also consider a staggered magnetic field $\Delta_{AF}$ in the $x{-}y$
plane: 
\begin{equation}\label{deltaxy}
{\cal H}_{AF} = \Delta_{AF} \sum_i (-1)^{R_i}
(c_{i,\uparrow}^{\dagger}c_{i,\downarrow} + 
c_{i,\downarrow}^{\dagger}c_{i,\uparrow}),
\end{equation}
where $\Delta_{AF}$ is a variational parameter that, together with the
chemical potential $\mu$ and the next-nearest-neighbor hopping of
Eq.~(\ref{meanfield}), can be determined by minimizing the variational energy
of ${\cal H}$. This kind of mean-field wave function was first introduced by 
Bouchaud, Georges, and Lhuillier~\cite{bouchaud} and then used to study 
${\rm He^3}$ systems and small atoms and molecules.~\cite{schmidt,schmidt2}
Recently, it has been also used to study the $t{-}J$ model on the 
triangular lattice.~\cite{weber} However, in these approaches the role of the 
long-range spin Jastrow factor was missed.
We emphasize that, in the mean-field Hamiltonian~(\ref{meanfield}), the 
magnetic order parameter is in the $x{-}y$ plane and not along the $z$ 
direction like:
\begin{equation}\label{deltaz}
{\cal H}_{AF} = \Delta_{AF} \sum_i (-1)^{R_i}
(c_{i,\uparrow}^{\dagger}c_{i,\uparrow} -
c_{i,\downarrow}^{\dagger}c_{i,\downarrow}).
\end{equation} 
Indeed, as already mentioned in the introduction, only in the case of 
Eq.~(\ref{deltaxy}) the presence of the spin Jastrow factor~(\ref{spinjastrow})
can introduce relevant fluctuations over the mean-field order parameter 
$\Delta_{AF}$, leading to an accurate description of the spin properties. 
By contrast, if the Jastrow potential is applied to the mean-field 
ansatz~(\ref{deltaz}), it cannot induce correct spin fluctuations and it is
not efficient in lowering the energy.

Finally, as already shown in Ref.~\onlinecite{dagotto}, the presence of the 
density Jastrow factor helps to reproduce the charge correlations of
the superconducting regime, giving rise to the correct Goldstone modes.

\begin{figure}
\includegraphics[width=0.50\textwidth]{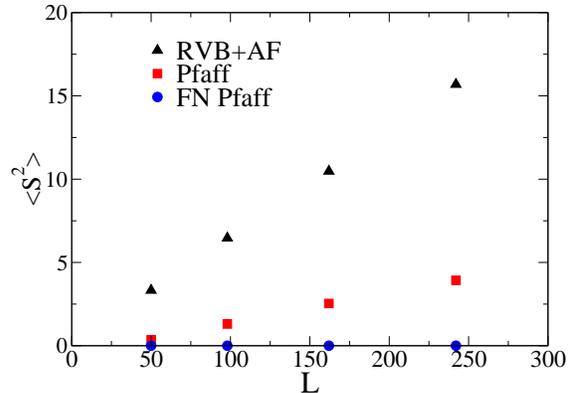}
\caption{\label{fig:s2}
Results for the total spin $\langle S^2 \rangle$ at half filling as a function
of the cluster size $L$ for the wave function of Eq.~(\ref{wf}) defined by 
the mean-field Hamiltonian~(\ref{meanfield}) and the two possible orientations 
of the magnetic field, i.e., Eqs.~(\ref{deltaxy}), indicated by ``Pfaff'',
and~(\ref{deltaz}), indicated by ``RVB+AF''. The FN results for the former
case are also shown.}
\end{figure} 

\begin{figure}
\includegraphics[width=0.50\textwidth]{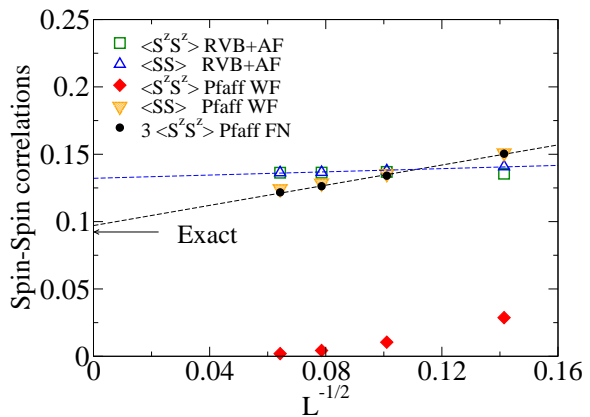}
\caption{\label{fig:mag}
Spin-spin correlations at the maximum distance at half filling 
for the wave functions of Fig.~\ref{fig:s2}. The exact value in the
thermodynamic limit is marked by the arrow.}
\end{figure} 

\begin{figure}
\includegraphics[width=0.50\textwidth]{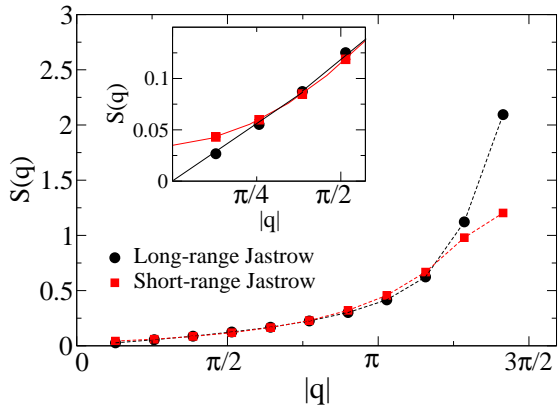}
\caption{\label{fig:sq}
Spin structure factor $S(q)$ at half filling for the variational wave function
of Eq.~(\ref{wf}) defined by the mean-field Hamiltonian of 
Eqs.~(\ref{meanfield}) and~(\ref{deltaxy}) with long-range and short-range 
(i.e., nearest-neighbor) Jastrow factors. Inset: Detail for small momenta.} 
\end{figure} 

The mean-filed Hamiltonian~(\ref{meanfield}) is quadratic in the fermionic 
operators and can be easily diagonalized in real space. Its ground state has 
the general form:
\begin{equation} \label{psimf}
|\Psi_{MF} \rangle = \exp  \left ( \frac{1}{2} \sum_{i,j,\sigma_i,\sigma_j} 
f^{\sigma_i,\sigma_j}_{i,j} c_{i,\sigma_i}^{\dagger} c_{j,\sigma_j}^{\dagger}
\right ) |0 \rangle,
\end{equation}
the pairing function $f^{\sigma_i \sigma_j}_{i j}$ being an antisymmetric 
$4L \times 4L$ matrix. Notice that in the case of the standard BCS 
Hamiltonian, with $\Delta_{AF}=0$ or even with $\Delta_{AF}$ along $z$, we 
have that $f^{\uparrow,\uparrow}_{i,j}=f^{\downarrow,\downarrow}_{i,j}=0$, 
while in presence of magnetic field in the $x{-}y$ plane the pairing function 
acquires non-zero contributions also in this triplet channel. 
The technical difficulty when dealing with such a state is that, given a 
generic configuration with definite $z$-component of the spin
$|x \rangle = c_{i_1,\sigma_1}^{\dagger} \dots c_{i_N,\sigma_N}^{\dagger}|0 \rangle$, 
we have that:
\begin{equation} \label{Pfaff}
\langle x|\Psi_{MF} \rangle = Pf[F],
\end{equation}
where $Pf[F]$ is the Pfaffian of the pairing function.~\cite{pfaffian} 
It should be noticed that, whenever 
$f^{\uparrow,\uparrow}_{i,j}=f^{\downarrow,\downarrow}_{i,j}=0$, the usual 
form of $\langle x|\Psi_{MF} \rangle$ written in terms of a determinant is
recovered. The fact of dealing with Pfaffians makes the algorithm slower than
the case of determinants, but the important point is that the algebra
of Pfaffians still allows us to have a very efficient updating procedure 
in the Monte Carlo calculation.
Then, by using the minimization technique described in 
Ref.~\onlinecite{sorella2}, we are able to deal with a large number of 
variational parameters and in particular we can optimize all the independent 
coefficient $v_{ij}$ and $u_{ij}$, beside the parameters contained in the 
mean-field Hamiltonian~(\ref{meanfield}).

\begin{figure}
\includegraphics[width=0.50\textwidth]{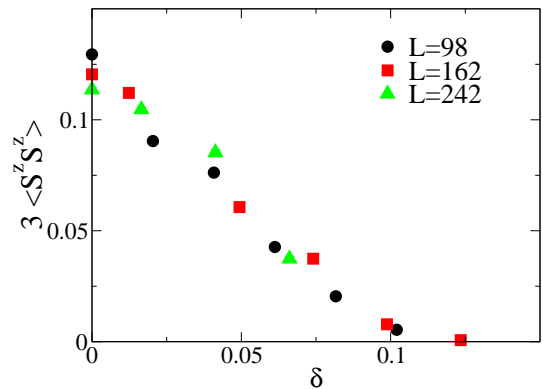}
\caption{\label{fig:magvsdop}
FN results for the spin-spin correlations (along the $z$ direction) at the 
maximum distance as a function of the doping for different sizes of the
cluster.}
\end{figure} 

\begin{figure}
\includegraphics[width=0.50\textwidth]{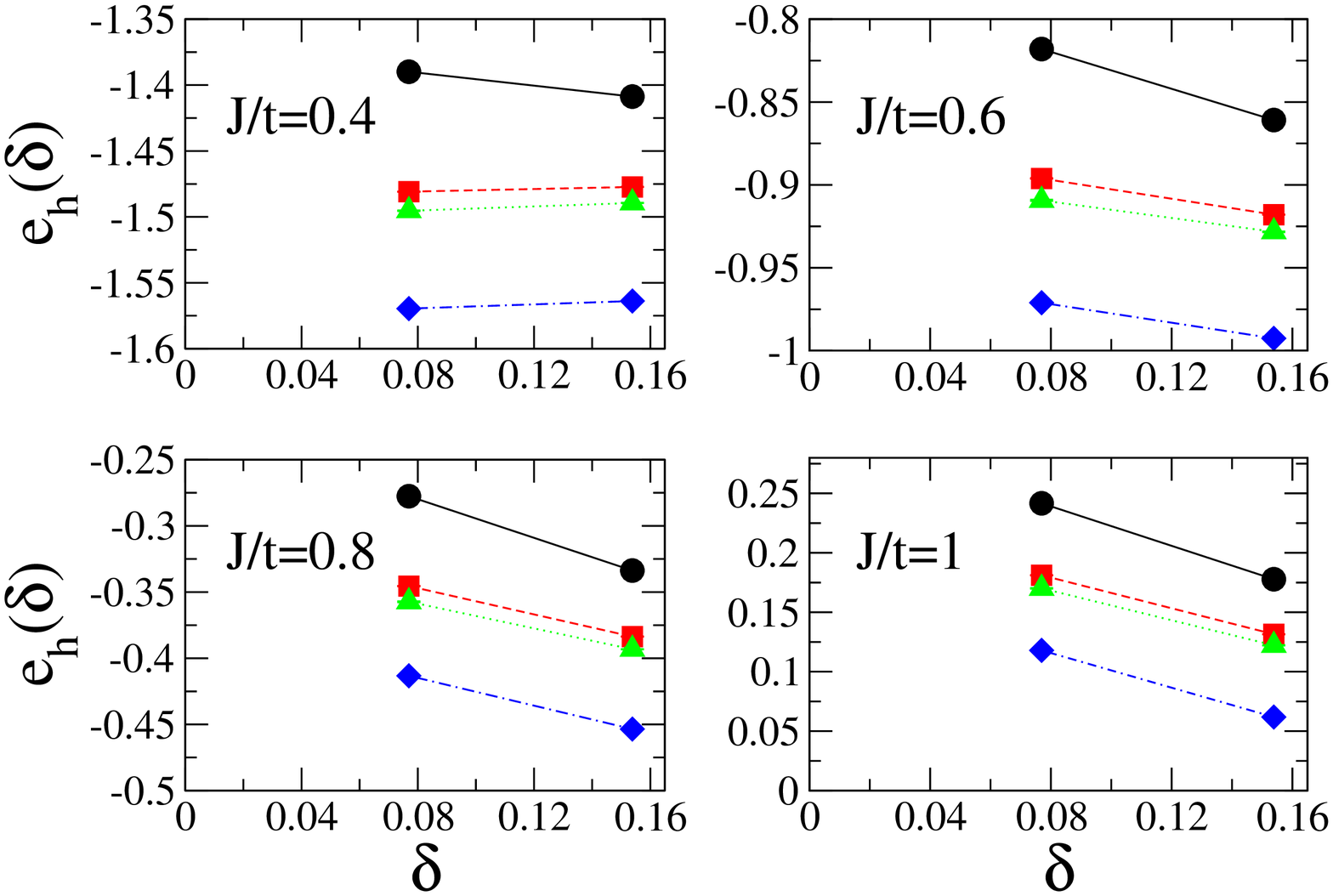}
\caption{\label{fig:slope26}
Energy per hole $e_h(\delta)$ as a function of the doping $\delta$ for
the 26-site cluster calculated by different approaches: The variational 
calculations for the Pfaffian wave function (circles), the FN approach of 
Eq.~(\ref{mixedav}) (squares), and the expectation value of the Hamiltonian 
over the FN ground state given by Eq.~(\ref{fnextrap}) (triangles); 
the exact results are also shown (diamonds).}
\end{figure} 

\begin{figure}
\includegraphics[width=0.50\textwidth]{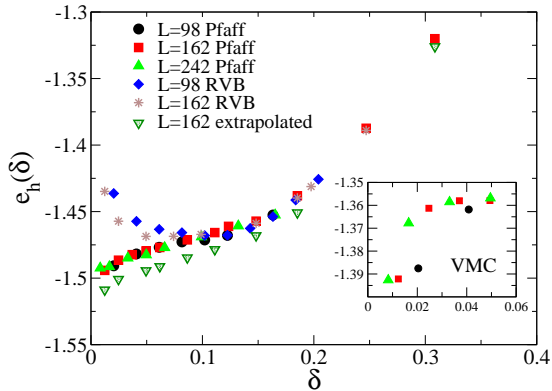}
\caption{\label{fig:emeryj04}
Energy per hole $e_h(\delta)$ as a function of the doping $\delta$ for 
$J/t=0.4$ and different sizes. The results are obtained by using the FN
approach described in the text. Two different states are used as guiding
function: The simple non-magnetic state, denoted by ``RVB'' and the
state with pairing, antiferromagnetism in the $x{-}y$ plane, and the spin
Jastrow factor, denoted by ``Pfaff''. The expectation value of the Hamiltonian 
over the FN ground state are also shown for $L=162$ for the latter case. 
Inset: Variational energy per hole for the Pfaffian wave function.}
\end{figure} 

\begin{figure}
\includegraphics[width=0.50\textwidth]{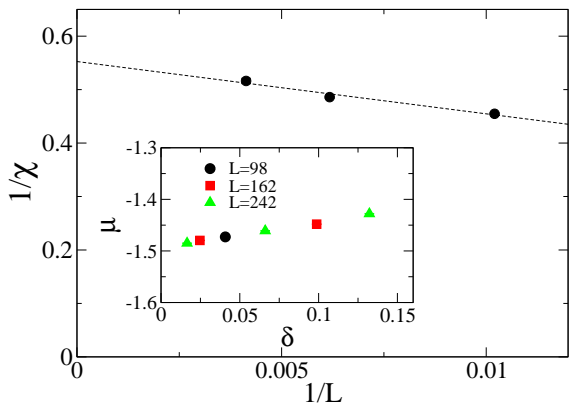}
\caption{\label{fig:comp}
The inverse compressibility of the half-filled Mott insulator for $J/t=0.4$
calculated by extracting the second derivative of the polynomial fit of the
FN energy. Inset: The chemical potential, defined through the difference of 
ground-state energies, as a function of the doping for different sizes of the 
cluster.}
\end{figure} 

The variational accuracy of a given wave function can be assessed by the FN
method that allows one to filter out the high-energy components of a given
state and to find the best variational state with the same nodes of the
starting one.~\cite{ceperley} On the lattice, the FN method can be simply 
defined as follows: Starting from the original Hamiltonian ${\cal H}$ we 
define an effective Hamiltonian by adding a perturbation $O$:
\begin{equation}\label{fnham}
{\cal H}_{eff}^\gamma = {\cal H} + (1+\gamma) O,
\end{equation}
here we follow Ref.~\onlinecite{sorella3} and introduce the external 
parameter $\gamma$, the original FN approximation of Ref.~\onlinecite{ceperley}
being recovered for $\gamma=0$.
The operator $O$ is defined through its matrix elements and depends upon
a given guiding function $|\Psi \rangle$, that is for instance the variational 
state itself, i.e., $|\Psi_{VMC} \rangle$:
\begin{equation}
O_{x^\prime,x} = \left \{
\begin{array}{ll}
-{\cal H}_{x^\prime,x} & {\rm if} \; s_{x^\prime,x} = \Psi_{x^\prime} {\cal H}_{x^\prime,x} \Psi_x >0 \nonumber \\
\sum_{y,s_{y,x}>0} {\cal H}_{y,x} \frac{\Psi_y}{\Psi_x} & {\rm for} \; x^\prime=x,
\end{array}
\right .
\end{equation}
where $\Psi_x = \langle x|\Psi \rangle$.
Notice that the above operator annihilates the guiding function, namely 
$O |\Psi \rangle=0$. Therefore, whenever the guiding function is close 
to the exact ground state of ${\cal H}$ the perturbation $(1+\gamma) O$ 
is expected  to be small and the effective Hamiltonian becomes very close
to the original one.
 
Let us review the properties of the FN Hamiltonian.
Trivially, for $\gamma=-1$, ${\cal H}_{eff}^\gamma$ coincides 
with ${\cal H}$, as the perturbation vanishes. The most important property 
of this effective Hamiltonian is that for $\gamma \ge 0$ its ground state 
$|\Psi_0^\gamma \rangle$ can be efficiently computed by using the Green's 
function Monte Carlo technique,~\cite{nandini,calandra2} that allows one to 
sample the distribution 
$\Pi_x \propto \langle x|\Psi \rangle \langle x|\Psi_0^\gamma \rangle$ 
by means of a statistical implementation of the 
power method: 
$\Pi \propto \lim_{n \to \infty } G^n \Pi^0$, where 
$\Pi^0$ is a starting distribution and $G_{x^\prime,x}= \Psi_{x^\prime} 
(\Lambda \delta_{x^\prime,x} - {\cal H}_{eff,x^\prime,x}^\gamma)/ \Psi_x$, 
is the so-called Green's function, defined with a large or even 
infinite~\cite{caprio} positive constant $\Lambda$, $\delta_{x^\prime,x}$
being the Kronecker symbol. The statistical method is very efficient for 
$\gamma \ge 0$, since in this case  all the matrix elements of $G$
are non-negative and, therefore, it can represent a transition 
probability in configuration space, apart for a normalization factor 
$b_x= \sum_{x^\prime} G_{x^\prime,x}$. 
In this case, it follows immediately that the asymptotic distribution $\Pi$ 
is also positive and, therefore, we arrive at the important conclusion that 
for $\gamma \ge 0$ the ground state of ${\cal H}_{eff}^\gamma$ has the same 
signs of the chosen guiding function. 
Within the FN approximation, we have a direct access to the ground-state 
energy $E_{FN}^\gamma$ of the effective Hamiltonian by sampling the 
so-called local energy 
$e_L(x) = \langle x| {\cal H} |\Psi \rangle / \langle x | \Psi \rangle$ over 
the distribution $\Pi_x$.
In the following, we will denote the standard FN energy for $\gamma=0$ simply 
by $E_{FN}$. It should be noted that, since $O |\Psi \rangle=0$, we have
that $E_{FN}^\gamma$ is also the mixed average of the original Hamiltonian 
\begin{equation}\label{mixedav}
E_{FN}^\gamma = 
\frac{\langle \Psi_0^\gamma| {\cal H}_{eff}^\gamma |\Psi_0^\gamma \rangle} 
{\langle \Psi_0^\gamma| \Psi_0^\gamma \rangle} =
\frac{\langle \Psi| {\cal H} |\Psi_0^\gamma \rangle}
{\langle \Psi| \Psi_0^\gamma \rangle}.
\end{equation}

$E_{FN}^\gamma$ gives a rigorous upper bound of the exact ground-state
energy $E_0=E_{FN}^{\gamma=-1}$ since it is an increasing function of $\gamma$
as the operator $O$ is positive definite~\cite{ceperleynote} and by the 
Hellman-Feynman theorem: 
\begin{equation}\label{hf}
\frac { d E_{FN}^\gamma} { d \gamma} =  
\frac{d \langle {\cal H}_{eff}^\gamma \rangle}{d \gamma} = 
\langle \frac{d {\cal H}_{eff}^\gamma}{d \gamma} \rangle =
\langle O \rangle \ge 0,
\end{equation}
here $\langle \dots \rangle$ indicates the expectation value over
$|\Psi_0^\gamma \rangle$.
This upper bound is also certainly below or equal to the variational energy
of the guiding function 
$E=\langle \Psi | {\cal H} |\Psi \rangle / \langle \Psi |\Psi \rangle$,
since from $O |\Psi \rangle =0$ it follows that $E$ is also the 
expectation value of the FN Hamiltonian over $|\Psi \rangle$, namely
$E=\langle \Psi | {\cal H}_{eff}^\gamma |\Psi \rangle / 
\langle \Psi |\Psi \rangle$.

One of the advantages of having introduced the parameter $\gamma$ is that it 
is possible to extract the expectation value of the original Hamiltonian 
${\cal H}$ over the FN state $|\Psi_0^\gamma \rangle$. Indeed, by   
applying Eq.~(\ref{hf}), we have that:
\begin{eqnarray}
E_{\Psi_0}^\gamma =\langle {\cal H} \rangle &=& 
\langle {\cal H}_{eff}^\gamma \rangle -(1+\gamma)
\frac{d \langle {\cal H}_{eff}^\gamma \rangle}{d \gamma} \nonumber \\
&=& E_{FN}^\gamma -(1+\gamma) \frac{d E_{FN}^\gamma}{d \gamma},
\label{extrapolation}
\end{eqnarray}
and therefore, by doing simulations for different values of $\gamma$ to
calculate numerically the derivative, it is possible to evaluate the
expectation value of ${\cal H}$ over the ground state of the FN Hamiltonian.
Moreover, by using the definition~(\ref{extrapolation}) and the fact that 
$E_{FN}^\gamma$ is a convex function,~\cite{sorella3} it turns out that:
\begin{equation}
\frac{d E_{\Psi_0}^\gamma}{d \gamma} = -(1+\gamma) 
\frac{d^2 E_{FN}^\gamma}{d \gamma^2} > 0,
\end{equation}
namely $E_{\Psi_0}^\gamma $ is monotonically increasing with $\gamma$.
A practical estimate of $E_{\Psi_0}^{\gamma=0}$, the best variational energy 
that can be obtained within a stable statistical method, can be worked out by 
performing two calculations for $\gamma=0$ and $\gamma =\tilde \gamma>0$ via:
\begin{equation}\label{fnextrap}
{\tilde E}_{\Psi_0}^{\gamma=0} = 
E_{FN} - \frac{1}{\tilde \gamma} (E_{FN}^{\tilde \gamma}-E_{FN}),
\end{equation}
${\tilde E}_{\Psi_0}^{\gamma=0}$ certainly improves the standard FN upper bound
of the energy and still 
${\tilde E}_{\Psi_0}^{\gamma=0} \ge E_{\Psi_0}^{\gamma=0}$.
This latter inequality follows from the convexity of 
$E_{FN}^{\gamma}$, implying that its first derivative at $\gamma=0$ is 
certainly larger or equal than the corresponding finite difference estimate. 
In order to obtain a compromise between having small enough statistical errors 
and a reasonable energy gain with respect to the mixed average of 
Eq.~(\ref{mixedav}), we have computed ${\tilde E}_{\Psi_0}^{\gamma=0}$ using 
$\tilde \gamma=1$.

\begin{figure}
\includegraphics[width=0.50\textwidth]{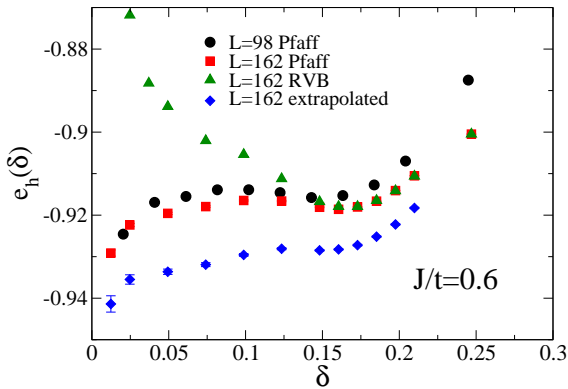}
\caption{\label{fig:emeryj06}
The same as in Fig.~\ref{fig:emeryj04} for $J/t=0.6$.}
\end{figure} 

\begin{figure}
\vspace{0.2cm}
\includegraphics[width=0.50\textwidth]{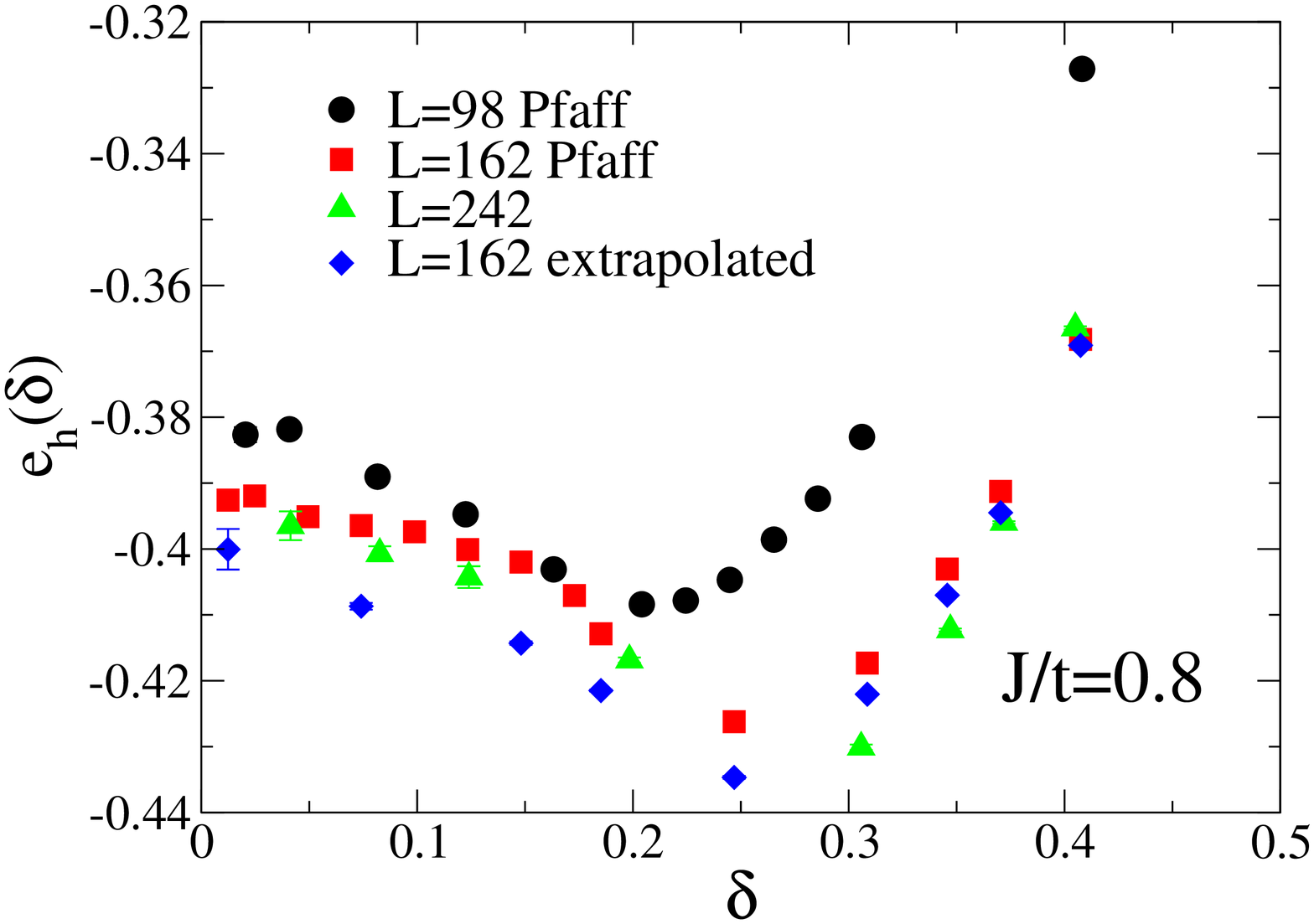}
\caption{\label{fig:emeryj08}
The same as in Fig.~\ref{fig:emeryj04} for $J/t=0.8$.}
\end{figure} 

In order to show the accuracy of the wave function~(\ref{wf}) and the FN
method, we report in Table~\ref{tab1} and~\ref{tab2} the energies 
for $2$ and $4$ holes in $26$ sites compared with the exact diagonalization 
data; in the same table we also show the results obtained from the 
wave function without the antiferromagnetic order parameter.
Finally, we report the values of the extrapolated energies 
${\tilde E}_{\Psi_0}^{\gamma=0}$ given by Eq.~(\ref{fnextrap}).
The inclusion of the magnetic field and the spin Jastrow 
factor strongly improve the energies with respect to the non-magnetic wave 
function. In particular, at half filling the FN is {\it exact} (within the 
error-bars), i.e., $E_{FN}/L=-1.184450(2)$ (in unit of $J=1$), whereas the 
variational energy is already very good $E_{VMC}/L=-1.18213(1)$. 
On the other hand, although the signs of the non-magnetic wave function
are correct (with the choice of $t_{i,j}$ and $\Delta_{i,j}$ connecting 
opposite sublattices and $\mu=0$), the non-magnetic wave function vanishes 
on many relevant configurations. This implies that, due to the importance
sampling procedure described before, such configurations are never visited 
by the Markov process, leading to $E_{FN}/L=-1.1833(3)$, despite the fact 
that the variational energy is not so poor $E_{VMC}/L=-1.15334(1)$.
We also notice that in this case the FN is highly unstable and many walkers 
are needed to stabilize its convergence.

\begin{table*}
\caption{\label{tab1}
Ground state energy for $2$ holes on $26$ sites and
different values of $J/t$. Two wave function with and without $\Delta_{AF}$ 
are indicated with ``Pfaff'' and ``RVB'', respectively. The variational results
are indicated by VMC and the fixed-node ones by FN. In the last two columns we
report the extrapolated value of Eq.~(\ref{fnextrap}) with the Pfaffian 
wave function and exact results by Lanczos method, respectively.}
\begin{tabular}{ccccccc}
\hline 
$J/t$ & $E^{RVB}_{VMC}/L$ & $E^{RVB}_{FN}/L$ & $E_{VMC}^{Pfaff}/L$ & $E_{FN}^{Pfaff}/L$ & ${\tilde E}_{\Psi_0}^{\gamma=0}/L$ & $E_{ex}/L$  \\
\hline \hline
0.3 & -0.48334(1) & -0.49256(1) & -0.48476(1) & -0.49325(1) & -0.49445(2) & -0.50097 \\
0.4 & -0.57664(1) & -0.58625(1) & -0.57978(1) & -0.58770(1) & -0.58881(2) & -0.59452 \\
0.5 & -0.67045(1) & -0.68091(1) & -0.67568(1) & -0.68327(1) & -0.68434(3) & -0.68945 \\
0.6 & -0.76463(1) & -0.77645(1) & -0.77228(1) & -0.77960(1) & -0.78062(3) & -0.78537 \\
0.8 & -0.95410(1) & -0.96920(1) & -0.96706(1) & -0.97414(1) & -0.97505(3) & -0.97935 \\
1.0 & -1.14483(1) & -1.16385(1) & -1.16352(1) & -1.17052(1) & -1.17136(2) & -1.17538 \\
\hline \hline
\end{tabular}
\end{table*}

\begin{table*}
\caption{\label{tab2}
The same as in Table~\ref{tab1} but for $4$ holes on $26$ sites.}
\begin{tabular}{ccccccc}
\hline 
$J/t$ & $E^{RVB}_{VMC}/L$ & $E^{RVB}_{FN}/L$ & $E_{VMC}^{Pfaff}/L$ & $E_{FN}^{Pfaff}/L$ & ${\tilde E}_{\Psi_0}^{\gamma=0}/L$ & $E_{ex}/L$  \\
\hline \hline
0.3 & -0.61372(1) & -0.62752(1) & -0.61478(1) & -0.62754(1) & -0.62958(3) & -0.64262 \\
0.4 & -0.68894(1) & -0.70101(1) & -0.68946(1) & -0.70106(1) & -0.70292(2) & -0.71437 \\
0.5 & -0.76461(1) & -0.77571(1) & -0.76512(1) & -0.77595(1) & -0.77770(4) & -0.78812 \\
0.6 & -0.84065(1) & -0.85132(1) & -0.84170(1) & -0.85189(1) & -0.85348(3) & -0.86337 \\
0.8 & -0.99361(1) & -1.00476(1) & -0.99709(1) & -1.00659(1) & -1.00806(2) & -1.01733 \\
1.0 & -1.14760(1) & -1.16072(1) & -1.15479(1) & -1.16422(1) & -1.16566(3) & -1.17493 \\
\hline \hline
\end{tabular}
\end{table*}

It is important to stress that the concomitant presence of the magnetic
order parameter $\Delta_{AF}$, that breaks the $SU(2)$ spin symmetry of the
electronic part, and the spin Jastrow factor of Eq.~(\ref{spinjastrow}), 
that also breaks the spin symmetry, gives rise to an almost symmetric state,
even for large sizes. This can be verified by calculating the total spin 
$S^2$: In Fig.~\ref{fig:s2} we report the results for the two wave functions 
with magnetic order in the $x{-}y$ plane and along the $z$ direction,
usually considered to describe the lightly doped region.~\cite{ivanov,gros2} 
In the same figure, we also report the FN value of $S^2$ (by using the former
state as the guiding function) in order to show that a totally 
symmetric state is eventually recovered. 

By a direct calculation of the spin-spin correlations at the maximum distance,
we obtain that also the value of the magnetization at half filling is in a
very good agreement with the exact result,~\cite{calandra2,sandvik} 
see Fig.~\ref{fig:mag}. It should be noted that the variational 
wave function with the magnetization in the $x{-}y$ plane and the spin 
Jastrow factor has very accurate isotropic spin-spin correlations, though 
in the $z$ direction they decay to zero in the thermodynamic limit. 
By performing the FN approach (with $\gamma=0$), a finite value for the 
correlations along $z$ is recovered. By contrast, when the magnetization is 
directed along $z$ in the variational ansatz, the spin correlations are 
almost Ising-like in the same direction and lead to overestimate the 
thermodynamic value of the magnetization, namely $m \sim 0.37$, 
see Fig.~\ref{fig:mag}.

Finally, we want to stress that the long-range tail of the spin Jastrow 
factor, obtained by minimizing the energy and leading to $v_q \sim 1/|q|$
for small $|q|$ ($v_q$ being the Fourier transform of $v_{ij}$), is necessary 
to correctly reproduce the small-$q$ behavior of the spin-structure factor
\begin{equation}
S(q)=\frac{1}{L} \sum_{l,m} e^{iq(R_l-R_m)} S^z_l S^z_m.
\end{equation}
Indeed, as it is clear from Fig.~\ref{fig:sq}, only with a long-range 
spin Jastrow factor, it is possible to obtain $S(q) \sim |q|$ for small momenta
and, therefore, a gapless spin spectrum. By contrast, with a short-range
spin Jastrow term (for instance with a nearest-neighbor term), 
$S(q) \sim {\rm const}$, for small $q$, that is clearly not correct.

\section{Results}\label{results}

Before considering the PS instability, we show in Fig.~\ref{fig:magvsdop}
the results for the spin-spin correlations at the maximum distance as
a function of the doping. We have that the magnetic order survives up to
$\delta \sim 0.1$, in agreement with previous 
calculations~\cite{ivanov,calandra3} and showing the importance to include
the magnetic parameter into the variational wave function.
Unfortunately, a precise size scaling analysis is not possible at finite 
hole concentration, since only discrete values of the doping are achievable 
and very rarely they are compatible from cluster to cluster.
 
Let us move to the central issue of this work.
In order to detect a possible PS instability, it is convenient to follow
the criterion given in Ref.~\onlinecite{lin} and consider the energy per hole:
\begin{equation}\label{eh}
e_h(\delta)=\frac{e(\delta)-e(0)}{\delta},
\end{equation}
where $e(\delta)$ is the energy per site at hole doping $\delta$ and $e(0)$ is
its value at half filling. For a stable system, $e_h(\delta)$ must be a
monotonically increasing function of $\delta$, since in this case the energy
is a convex function of the doping and $e_h(\delta)$ represents the chord 
joining half filling and the doping $\delta$. On the other hand,
the PS instability is marked by a minimum at a given $\delta_c$ on finite 
clusters, and a flat behavior up to $\delta_c$ in the thermodynamic limit 
where the Maxwell construction is implied.

Firstly, Fig.~\ref{fig:slope26} shows the results of $e_h(\delta)$
for different ratios $J/t$ on the 26-site cluster, where the exact data are 
available by the Lanczos method. Although these data are already contained 
in tables~\ref{tab1} and~\ref{tab2}, their graphical representation better
shows our accuracy to estimate the slope of the energy per hole.
In particular, we stress the fact that, even though already the variational 
results of the wave function~(\ref{wf}) are very accurate, there is a strong 
improvement by considering the FN approach, both in the mixed average of 
Eq.~(\ref{mixedav}) and in the extrapolation of Eq.~(\ref{fnextrap}), for
which a perfect estimation of the slope is obtained.

Then we can move to large cluster to extract the thermodynamic properties.
We report in Fig.~\ref{fig:emeryj04} the results of the energy per hole for
$J/t=0.4$. For comparison, the FN calculations for $\gamma=0$ are performed 
by using two different guiding functions, including or not the 
antiferromagnetic order parameter and the spin Jastrow factor. 
At large doping the results are independent on the choice of the guiding 
state, clearly indicating that the antiferromagnetism is not essential. 
However, by decreasing the hole concentration, the inclusion of the 
antiferromagnetic order becomes crucial for the stabilization of the 
homogeneous phase, whereas the simple projected BCS state is eventually 
unstable at small doping. This latter outcome actually is in agreement with 
our previous calculations~\cite{calandra} and confirms what has been noticed 
by Hellberg and Manousakis~\cite{manousakis2} and interpreted as an evidence 
for PS close to the insulating limit.
By contrast, our present FN results, based on the wave function with 
antiferromagnetic fluctuations, strongly improve
the accuracy of previous calculations for small doping and point
towards the stability of the homogeneous phase for all hole concentrations.
Quite impressively, the energies are very accurate on the whole doping regime
analyzed and there is not a qualitative difference if one considers the 
expectation value of the Hamiltonian~(\ref{fnextrap}),
see Fig~\ref{fig:emeryj04}. These results indicate that the ground state is
stable for all the hole concentrations, namely down to $\delta \sim 0.01$ 
(i.e., two holes on 242 sites), strongly improving our previous estimate of 
the phase diagram. Remarkably, also the variational wave function is stable 
for such value of the super-exchange interaction and small hole concentrations, 
see the inset of Fig~\ref{fig:emeryj04}.
To our knowledge, this is the first successful attempt to obtain a variational 
state which is clearly stable towards the formation of regions with segregated
holes, when approaching the Mott insulating regime.

\begin{figure}
\includegraphics[width=0.50\textwidth]{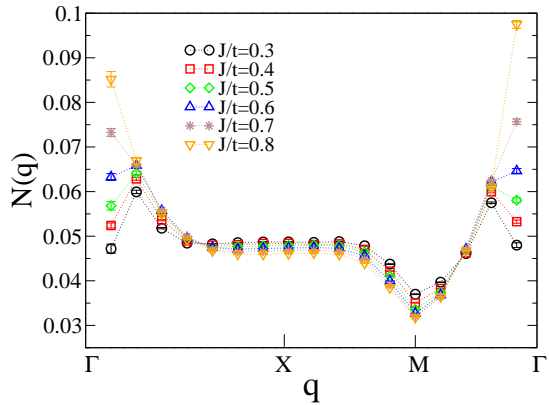}
\caption{\label{fig:nq8holes}
FN results for the density correlation function for $8$ holes on $162$ sites 
and different values of $J/t$. The high-symmetry points are marked as 
$\Gamma=(0,0)$, $X=(\pi,\pi)$, and $M=(\pi,0)$.}
\end{figure} 

\begin{figure}
\includegraphics[width=0.50\textwidth]{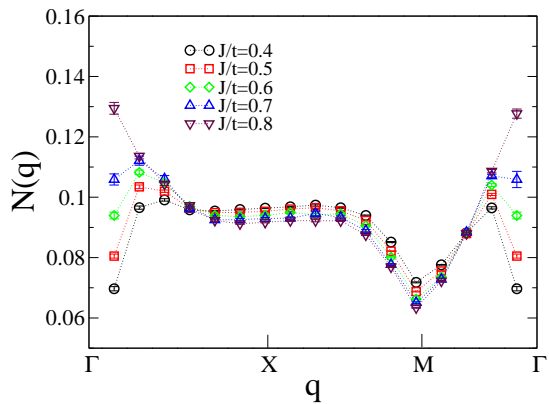}
\caption{\label{fig:nq16holes}
The same as in Fig.~\ref{fig:nq8holes} but for $16$ holes on $162$ sites.}
\end{figure} 

From the energy calculation it is straightforward to estimate the 
compressibility $\chi$ for $\delta \to 0$:
\begin{equation}
\chi^{-1} = \frac{\partial^{2} e(\delta)}{\partial \delta^{2}}.
\end{equation}
Recently, Imada and coworkers,~\cite{imada,assaad} by using hyper-scaling 
arguments and numerical simulations on the Hubbard model, proposed that the 
compressibility must diverge when the insulating phase is approached 
by decreasing the doping concentration. Their arguments imply that 
$e(\delta) \sim \delta^3$ for small doping, as in the one-dimensional case, 
where the charge properties can be simply understood by considering spinless 
fermions. Instead, within our FN approach, we find that the compressibility 
stays finite up to half filling. Indeed, for $J/t=0.4$ and in general for the 
stable magnetic phase, the variational calculation provides a finite 
compressibility that is further decreased by the more accurate FN 
approximation. It should be noticed that a much larger
compressibility, or even an infinite one, could be worked out when considering
only small size calculations, like the ones used in Ref.~\onlinecite{imada} 
to obtain $\chi \sim |\mu - \mu_c|^{-1/2} \sim \delta^{-1}$ (where $\mu$ is
the chemical potential and $\mu_c$ is nothing but the charge gap at half 
filling): In this case, it is possible to underestimate the slope of the 
energy at small doping and, therefore, also to overestimate the value of 
$\chi$. Instead, from our large cluster calculations, we have a 
clear evidence that the chemical potential is linear with the doping close 
to half filling or, equivalently, that $e(\delta) \sim \delta^2$, implying 
a finite compressibility when $\delta \to 0$, see Fig.~\ref{fig:comp}. 
Our calculations are rather robust and do not depend upon the number of holes 
considered and a very accurate polynomial fit of the energy turns out to be 
very stable. We argue that the infinite compressibility scenario proposed by 
Imada and coworkers could be correct when the antiferromagnetism 
does not play an important role and the undoped system is a spin liquid with 
no magnetic order. This is also supported by dynamical mean-field theory 
calculations by Kotliar and coworkers~\cite{kotliar} on the Hubbard model,
where the mean-field solution without an antiferromagnetic order parameter
leads to a diverging compressibility close to the Mott regime.

By increasing the antiferromagnetic super-exchange, we come closer to the PS 
region. Indeed, for $J/t=0.6$ we obtain that the energy per hole $e_h(\delta)$ 
shows a slightly non-monotonic behavior with a minimum for 
$\delta_c \sim 0.17$, when considering the FN energies. This minimum 
disappears by performing the extrapolation of Eq.~(\ref{fnextrap}) to 
estimate the expectation value of the $t{-}J$ Hamiltonian over the FN ground 
state, see Fig.~\ref{fig:emeryj06}. This fact would indicate that, for this 
value of $J/t$, the FN Hamiltonian~(\ref{fnham}) has an higher tendency 
towards PS than the original $t{-}J$ model.
In this case, the mixed average of Eq.~(\ref{mixedav}) is slightly biased,
and this bias can be eliminated by considering the 
actual expectation value of the $t{-}J$ Hamiltonian over the FN ground state. 
In doing this, we approach the exact result (by improving the energy) and an 
homogeneous phase, with a monotonically increasing energy per hole, is 
obtained. Within this more accurate scheme, we substantially improve 
our previous results that were based on the mixed average of the FN 
approximation and that indicated a rather high critical doping.~\cite{calandra}
Unfortunately, within our numerical approach, it is very difficult to 
study the possible formation of hole droplets close to the PS instability, 
as suggested by Poilblanc.~\cite{poilblanc}
Indeed, this would require a very delicate size scaling of the binding 
energy of few holes, which is beyond our present possibilities.

By further increasing the super-exchange coupling, we eventually enter into
the PS region: For $J/t=0.8$, the energy per hole has a rather deep minimum
at finite doping and also the expectation value~(\ref{fnextrap}) clearly
indicates a non-monotonic behavior, see Fig.~\ref{fig:emeryj08}. 

\begin{figure}
\includegraphics[width=0.50\textwidth]{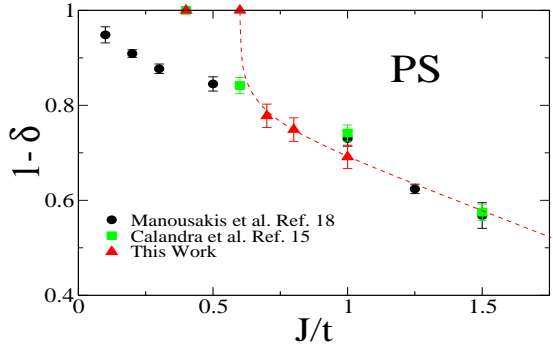}
\caption{\label{fig:phase}
Boundary for the phase separation (PS) instability. The results of previous
works are also shown for comparison. The line is a guide to the eye.}
\end{figure} 

Finally, it is important to stress that very similar results can be also 
obtained by considering the density-density correlation function
\begin{equation}
N(q)=\frac{1}{L}\sum_{l,m} e^{iq(R_l-R_m)} n_l n_m.
\end{equation}
In this case, since $N(q)$ is a diagonal operator in the configuration space, 
it is easy to compute its average value over the FN ground state by using 
the so-called forward-walking technique.~\cite{calandra2} This quantity is
therefore free from possible bias coming from mixed averages.
The PS instability is signaled by the divergence at small momenta of $N(q)$. 
In our previous paper,~\cite{calandra} we reported the calculations of this 
quantity, showing the presence of a finite-$q$ peak, linearly depending upon 
the doping, close to the PS instability. Here, thanks to the accuracy of the 
guiding function and the progress in stabilizing the statistical 
implementation of the FN technique, we are able to present much more accurate 
results that confirm the previous ones. Indeed, the existence of this peak 
is due to the closeness of the PS: Figs.~\ref{fig:nq8holes} 
and~\ref{fig:nq16holes} show the evolution of $N(q)$ by 
increasing $J/t$ for two values of the doping, near the insulating regime.
In particular, we obtain the evidence for a stable homogeneous phase
for $J/t \sim 0.4$, confirming the indications given by the analysis based
upon the energy per hole. Then also the progressive development of a huge 
peak around $q=(0,0)$ for $J/t \sim 0.7$ is in good agreement with the
energy calculations. All together, these results allow us to draw our final 
phase diagram of Fig.~\ref{fig:phase}, where we report, for comparison,
also some of the previous estimations for the PS boundaries.

\section{Conclusion}\label{conclusions}

We have revisited the problem of the PS instability in the $t{-}J$ model.
By generalizing the Pfaffian wave function introduced some time 
ago,~\cite{bouchaud} we have defined a very accurate variational state that, 
for the first one to our knowledge, is stable against PS at low doping. 
In particular, we have shown the necessity to consider both an 
antiferromagnetic order parameter (in the fermionic determinant) and a spin
Jastrow factor, to mimic the spin fluctuations. In this way all the 
low-energy properties of the exact ground state are correctly reproduced.
Then, by using a more sophisticated Monte Carlo technique that can filter out
the high-energy components of a given trial wave function, we can obtain
the ground state of an effective Hamiltonian and, at the same time, assess 
the stability our initial guess.
So, we have shown that for $J/t=0.4$, the ground state does not
phase separate at any hole doping down to $\delta \sim 0.01$, giving a 
serious improvement on the possible PS boundaries at small $J/t$.
Remarkably, the analysis based on the energy per hole is also corroborated 
by the calculation of the static density-density correlations.
The phase separation, in the low doping region, appears at a critical 
antiferromagnetic coupling slightly larger than the value given in
Ref.~\onlinecite{calandra}, namely here we find $J_c/t \sim 0.7$.
Although future improvements in the Monte Carlo technique or in the accuracy 
of the variational wave function may lead to an higher coupling, it looks 
unlikely to reach the critical point recently obtained by high-temperature 
expansion,\cite{putikka,putikka2} i.e., $J_c/t \sim 1.2$. 
In fact, as shown in Fig.~\ref{fig:slope26}, our present accuracy in the 
energy per hole is about $~0.05t$ and its slope is almost correct.
This holds rather independently of $J/t$ and system sizes, at least for the 
clusters where exact results are available. 
For $J/t=0.8$ (see Fig.~\ref{fig:emeryj08}), the minimum of the energy 
per hole implies an energy gain for the inhomogeneous phase of about $0.05t$ 
per hole, i.e., comparable with our maximum possible error estimated before.
Thus we expect that $J_c/t$ cannot be much larger than $0.8$ even for a 
numerically exact method.

Moreover, we have obtained that, in contrast with what was found in the
Hubbard model, the compressibility stays finite by approaching the Mott 
insulator. A simple explanation of a finite compressibility in two dimensions 
is obtained by by assuming that the holes form hole pockets around the 
nodal points [i.e., $q=(\pm \pi/2, \pm \pi/2)$] and behave as spinless 
fermions, implying that $e(\delta) \simeq \delta^{1+2/D}$, where 
$D$ is the spatial dimension. In this simple scenario the 
compressibility is divergent only in one dimension, whereas it is finite in 
two dimensions, and should approach zero in three dimensions, leading to a 
more conventional metal-insulator transition.

The stability against phase separation of a wave function with explicit 
antiferromagnetism and d-wave superconducting order parameter provides new 
insights for understanding the phase diagram of the high-temperature 
superconductors. Remarkably, in the clean system, possibly idealized by 
the $t{-}J$ model, the antiferromagnetism and the d-wave order parameter 
should not exclude each other, at least at the variational level, 
and actually cooperate to decrease the energy and lead to a stable homogeneous 
phase.

This research has been supported by PRIN-COFIN 2005 and CNR-INFM.


\begin{thebibliography}{99}
\bibitem{pfaffian} In general, the Pfaffian of a $2n \times 2n$ antisymmetric 
   matrix $\chi$ is equal to the antisymmetric product of its matrix 
   elements, i.e., 
   $Pf[\chi] = {\cal A} \{ \chi_{1,2} \chi_{3,4} \dots \chi_{2n-1,2n} \}$, 
   with the constraint that each term 
   $\chi_{i_1,j_1} \chi_{i_2,j_2} \dots \chi_{i_n,j_n}$ has $i_k<j_k$ and
   $i_1<i_2< \dots i_n$.
\bibitem{ceperleynote} This has been shown in Ref.~\onlinecite{ceperley}, 
   by proving that $\langle \Phi |O|\Phi \rangle \ge 0 $ for any wave function 
   $|\Phi \rangle$.
\end{thebibliography}
\end{document}